%% file: ms.tex
\documentclass[12pt]{article}
\usepackage{graphicx}
\usepackage{subcaption} 
\usepackage{amsmath,amsthm,amssymb,latexsym}

\newcommand\pubnumber{NuPhys2015-Li}
\newcommand\pubdate{\today}

\def\umn{School of Physics and Astronomy, University of Minnesota\\
116 Church Street S.E., Minneapolis, MN 55455, USA}

\def\Title#1{\begin{center} {\Large #1 } \end{center}}
\def\Author#1{\begin{center}{ \sc #1} \end{center}}
\def\Address#1{\begin{center}{ \it #1} \end{center}}

\newcommand\pubblock{\rightline{\begin{tabular}{l} \pubnumber\\
         \pubdate  \end{tabular}}}
\newenvironment{Abstract}{\begin{quotation}  }{\end{quotation}}
\newenvironment{Presented}{\begin{quotation} \begin{center} 
             PRESENTED AT\end{center}\bigskip 
      \begin{center}\begin{large}}{\end{large}\end{center} \end{quotation}}
\def\Acknowledgements{\bigskip  \bigskip \begin{center} \begin{large}
             \bf ACKNOWLEDGEMENTS \end{large}\end{center}}

\input econfmacros.tex

\begin{document}
\begin{titlepage}
\pubblock

\vfill
\Title{Do Neutrino Wave Functions Overlap and Does it Matter?}
\vfill
\Author{Cheng-Hsien Li \& Yong-Zhong Qian}
\Address{\umn}
\vfill
\begin{Abstract}
Studies of neutrinos commonly ignore anti-symmetrization 
of their wave functions. This implicitly assumes that either
spatial wave functions for neutrinos with approximately the 
same momentum do not overlap or their overlapping has
no measurable consequences. We examine these assumptions by 
considering the evolution of three-dimensional neutrino wave packets 
(WPs). We find that it is perfectly adequate to treat accelerator and 
reactor neutrinos as separate WPs for typical experimental setup.
While solar and supernova neutrinos correspond to overlapping WPs,
they can be treated effectively as non-overlapping for analyses of 
their detection.
\end{Abstract}
\vfill
\begin{Presented}
NuPhys2015, Prospects in Neutrino Physics\\
Barbican Centre, London, UK,  December 16--18, 2015
\end{Presented}
\vfill
\end{titlepage}
\def\thefootnote{\fnsymbol{footnote}}
\setcounter{footnote}{0}

\section{Introduction}
Quantum particles can be treated as wave packets 
(WPs). The use of neutrino WPs clarified some conceptual confusion 
\cite{Akhmedov2009} from the use of plane waves to discuss 
neutrino oscillations and introduced a few oscillation-suppressing 
terms in the flavor transformation probabilities. For that problem, 
one-dimensional (1D) consideration suffices as only the longitudinal 
evolution of neutrino WPs is of concern. Nevertheless, the simplified 
1D description is not a full account of WP propagation in the 3D space.

\begin{figure}[h]
\centering
  \begin{subfigure}[b]{0.45\textwidth}
  \centering
   \includegraphics[scale=0.35]{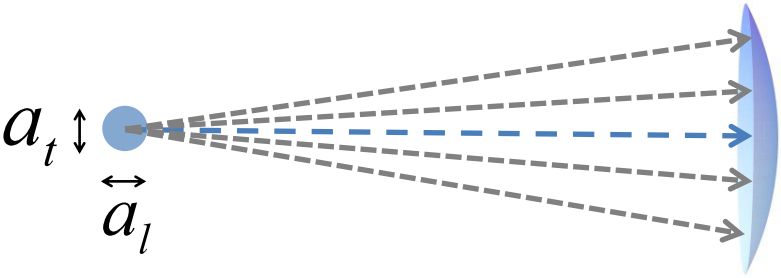}
     \vspace{0.3 in}
  \caption{}
  \label{fig:WP spread}
  \end{subfigure}
  \begin{subfigure}[b]{0.45\textwidth}
  \centering
 \includegraphics[scale=0.45]{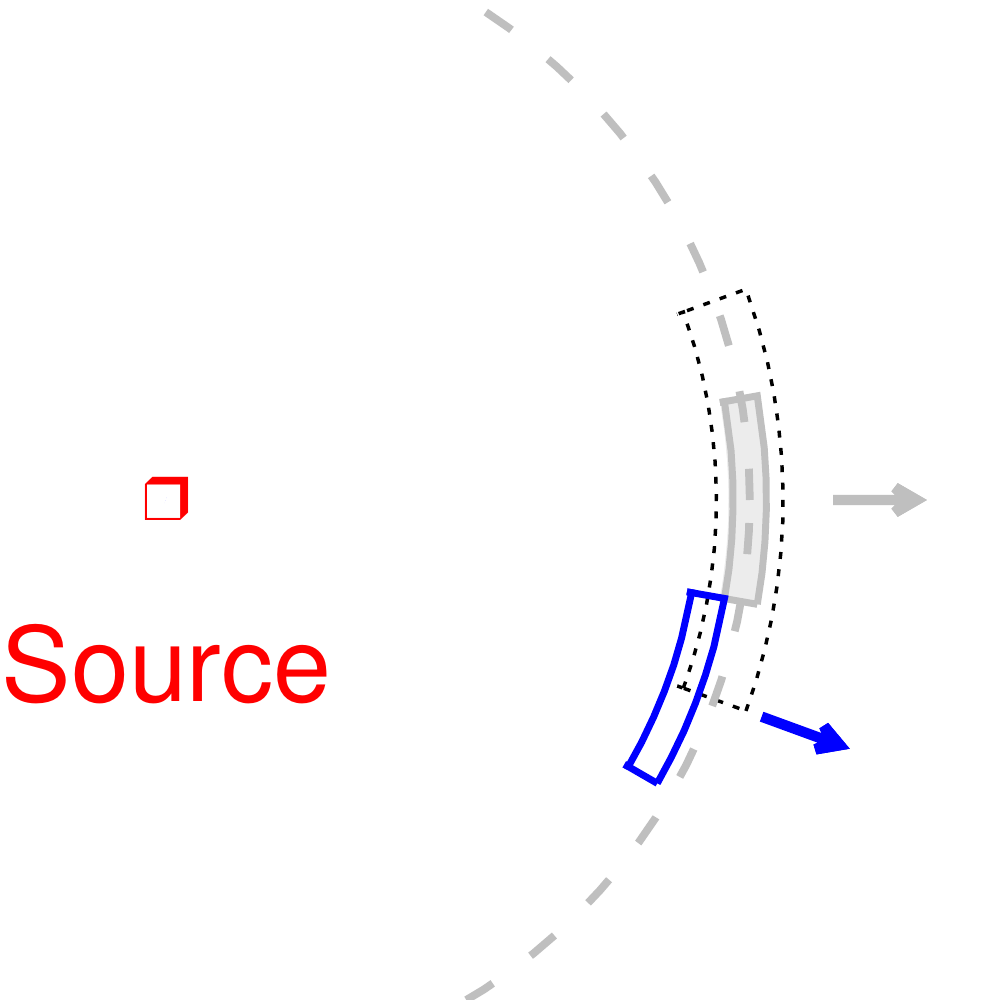}
  \caption{}
  \label{fig:Overlap criteria}
  \end{subfigure}
  \caption{(a) Spreading of a massless neutrino WP with 
  initial position uncertainties $a_l$ and $a_t$. The blue dashed 
  arrow shows the classical path. (b) Criteria for overlap. 
  The blue and gray strips represent the 90\%-probability 
  volumes of two WPs.}
  \label{fig:Illustration of WP spreading and overlap}
\end{figure}

According to Heisenberg's uncertainty principle, a particle localized 
in a finite spatial region has intrinsic momentum uncertainty that
causes the WP to spread over time. The longitudinal spreading of 
a neutrino WP is suppressed by the tiny neutrino mass \cite{Giunti1991} 
and can be neglected. In contrast, the transverse size of the WP 
increases with the time of travel $t$ as $(\Delta k_\perp / k_0)t$, where 
$\Delta k_\perp$ and $k_0$ are the transverse momentum uncertainty 
and the average momentum of the WP, respectively. Limited by the 
speed of light, the WP asymptotically evolves into a spherical shape 
that subtends a constant angle from its initial position, as depicted in 
Fig.~\ref{fig:WP spread}.

The transverse spreading potentially allows neutrino WPs emitted 
in slightly different directions but with approximately the same 
momentum to overlap. If such overlap occurs, the indistinguishability 
of neutrinos would require a formal many-particle treatment. 
Here we describe an approximate approach to quantify the overlap 
based on the evolution of 3D Gaussian WPs. We apply this approach
to accelerator, reactor, solar and supernova neutrinos and find no and 
severe overlap for the former and latter two sources, respectively.
However, the overlap has no measurable consequences for practical
detection of solar and supernova neutrinos.

\section{Estimating Overlap of 3D WPs}
As the tiny neutrino mass introduces negligible longitudinal WP
spreading, we assume massless neutrinos to focus on the transverse 
spreading. We consider a Gaussian WP with momenta closely
centered around the mean value $\vec{k}_0 = k_0\hat{z}$. An initial  
WP with the position widths shown in Fig.~\ref{fig:WP spread}
can be expressed as
\begin{equation}
\Psi(\vec{r},0)=\frac{1}{(2\pi)^{3/4}a_ta_l^{1/2}}
\exp\left(
-\frac{\rho^2}{4a_t^2}-\frac{z^2}{4a_l^2}
+ik_0z
\right),
\end{equation}
where $\rho \equiv \sqrt{x^2 + y^2} $. At time $t>0$,
the above WP evolves into \cite{Li2016}
\begin{equation}
\label{eq: Psi in the far-field limit at t>0 approximated result}
\Psi(\vec{r},t)\propto 
\frac{1}{z}\exp\left\{
			-\frac{\left[z+\rho^2/(2z)-t\right]^2}{4a_l^2}
			-\frac{\rho^2}{z^2(k_0a_t)^{-2}}
+ik_0\left(z+\frac{\rho^2}{2z}-t\right)
	\right\}.
\end{equation}
With $r \approx z + \rho^2/(2z)$, the corresponding probability 
density can be expressed in spherical coordinates 
as $|\Psi(\vec{r},t)|^2 = R(r,t)\Theta(\theta)$, where
\begin{equation}
\label{eq: radial and angular probability distribtuions }
R(r,t)\propto \frac{1}{r^2}
\,\exp\left[
-\frac{(r-t)^2}{2a_l^2}
\right],\ 
\Theta(\theta)\propto 
\exp\left[
-\frac{\theta^2}{2\cdot\left(2k_0a_t\right)^{-2}}
\right]. 
\end{equation}
For estimating the overlap of WPs, we define the 90\%-probability 
volume by $t-2a_l<r<t+2a_l$ and $0\leq\theta\lesssim 1.22(k_0 a_t)^{-1}$, 
which corresponds to the 95\%-probability regions of the radial and 
angular distributions.

We regard two WPs as overlapping if their 90\%-probability volumes 
intersect (see Fig.~\ref{fig:Overlap criteria}) and if their energies are 
the same within the intrinsic uncertainty $\Delta E$. This defines the 
emission time window $\tau\sim\mathcal{O}[(\Delta E)^{-1}]$ and solid 
angle $\Delta\Omega_\text{overlap}\sim\mathcal{O}[(k_0 a_t)^{-2}]$ 
(see Fig.~\ref{fig:Overlap criteria}) for the WPs of concern. 
With a differential production rate 
$d^2\Phi/dE_\nu d\Omega$ for the source, the number of WPs 
expected to overlap with a reference WP is 
\begin{equation}
\label{eq: overlap factor}
\eta=\tau\frac{d^2\Phi}{dE_\nu d\Omega}(\Delta E) 
(\Delta\Omega_\text{overlap})
\sim\frac{d^2\Phi}{dE_\nu d\Omega}
\frac{96\pi}{(k_0 a_t)^2}\sim 96\pi\frac{d^2\Phi}{dE_\nu d\Omega}
\left(\frac{\Delta k_\perp}{k_0}\right)^2,
\end{equation}
where $\Delta k_\perp\sim a_t^{-1}$ is the transverse momentum 
uncertainty of the WPs. Because the WPs have sharply-peaked 
momentum distributions, $\Delta k_\perp\ll k_0$ and
\begin{equation}
\eta\ll 96\pi\frac{d^2\Phi}{dE_\nu d\Omega}.
\label{eq:eta}
\end{equation}
The numerical factor in Eqs.~(\ref{eq: overlap factor}) and
(\ref{eq:eta}) comes from generous estimates of
$\Delta E$, $\tau$, and $\Delta\Omega_\text{overlap}$. We will
see that this factor does not affect our results below.

We give characteristic parameters for accelerator, reactor, solar, 
and supernova neutrinos along with the corresponding estimates 
of $\eta(k_0a_t)^2$ in Table~\ref{tab:overlap factor for sources}.
It can be seen that accelerator and reactor neutrinos can be safely 
treated as non-overlapping WPs. However, for reasonable guesses 
of $k_0a_t$, solar and supernova neutrinos correspond to
overlapping WPs. 

\begin{table}[t]
\begin{center}
\caption{Characteristics of various neutrino sources}
\begin{tabular}{lccccc}
\hline
\hline
Source & $d^2\Phi/dE_\nu d\Omega$&$\eta(k_0 a_t)^2$ & $D$&$E_\nu$& HBT\\
&(MeV$^{-1}$~s$^{-1}$~sr$^{-1}$)&&(m)&(MeV)&setup\\
\hline
accelerator &$\sim 10^{18}$&$\sim 0.1$  & $\sim10^6$& 
$\sim10^3$ & No \\
reactor  &$\sim 10^{19}$ &$\sim 1$ &  $\sim10^3$--$10^5$& 
$\sim 1$ & No \\
sun &$\sim 10^{32}$--$10^{38}$ &$\sim10^{13}$--$10^{19}$ & 
$\sim 10^{11}$& $\sim 0.1$--10 & No \\
supernova & $\sim 10^{51}$--$10^{55}$&$\sim 10^{32}$--$10^{36}$ & 
$\sim 10^{20}$&  $\sim 10$ & Yes\\
\hline
\hline
\end{tabular}
\label{tab:overlap factor for sources}
\end{center}
\end{table}
\vspace{-0.2 in}

\section{Physics of Quantum WPs: Overlapping or Not}
When WPs do not overlap, one might ask how
the 3D WP treatment can be reconciled with the picture 
of bullet-like particles, which is commonly assumed for
analyzing neutrino detection. As the transverse size of 
the WP can easily become macroscopically large, any 
microscopic detection process only ``sees'' a 1D neutrino 
wave train weighted by a direction-dependent amplitude, 
the square of which is $\Theta(\theta)$ in 
Eq.~\eqref{eq: radial and angular probability distribtuions }. 
Because of the spherical wave front in 
Eq.~\eqref{eq: Psi in the far-field limit at t>0 approximated result}, 
the observed effective plane-wave momentum points in the same 
direction as that defined in the classical sense. In addition, 
without interference among the WPs, the observed
flux is a sum over WPs emitted in different directions. As 
$\Theta(\theta)$ is normalized, this flux is the same as 
the particle number flux from a source emitting bullet-like particles. 

When WPs overlap, formally neutrinos should be described by 
an anti-symmetric many-particle wave function. However,
if among the overlapping WPs, only one neutrino is detected 
at a time, it can be shown that the interference terms in the 
one-particle detection probability are proportional to the inner 
products of the one-particle states. Although these states overlap 
in both position and momentum spaces at the detector, they are 
in fact orthogonal because their production processes are spatially 
separated at the source. Consequently, the one-particle detection 
rate is not affected by the overlap of WPs.

\begin{figure}[t]
\centering
\includegraphics[scale=0.20]{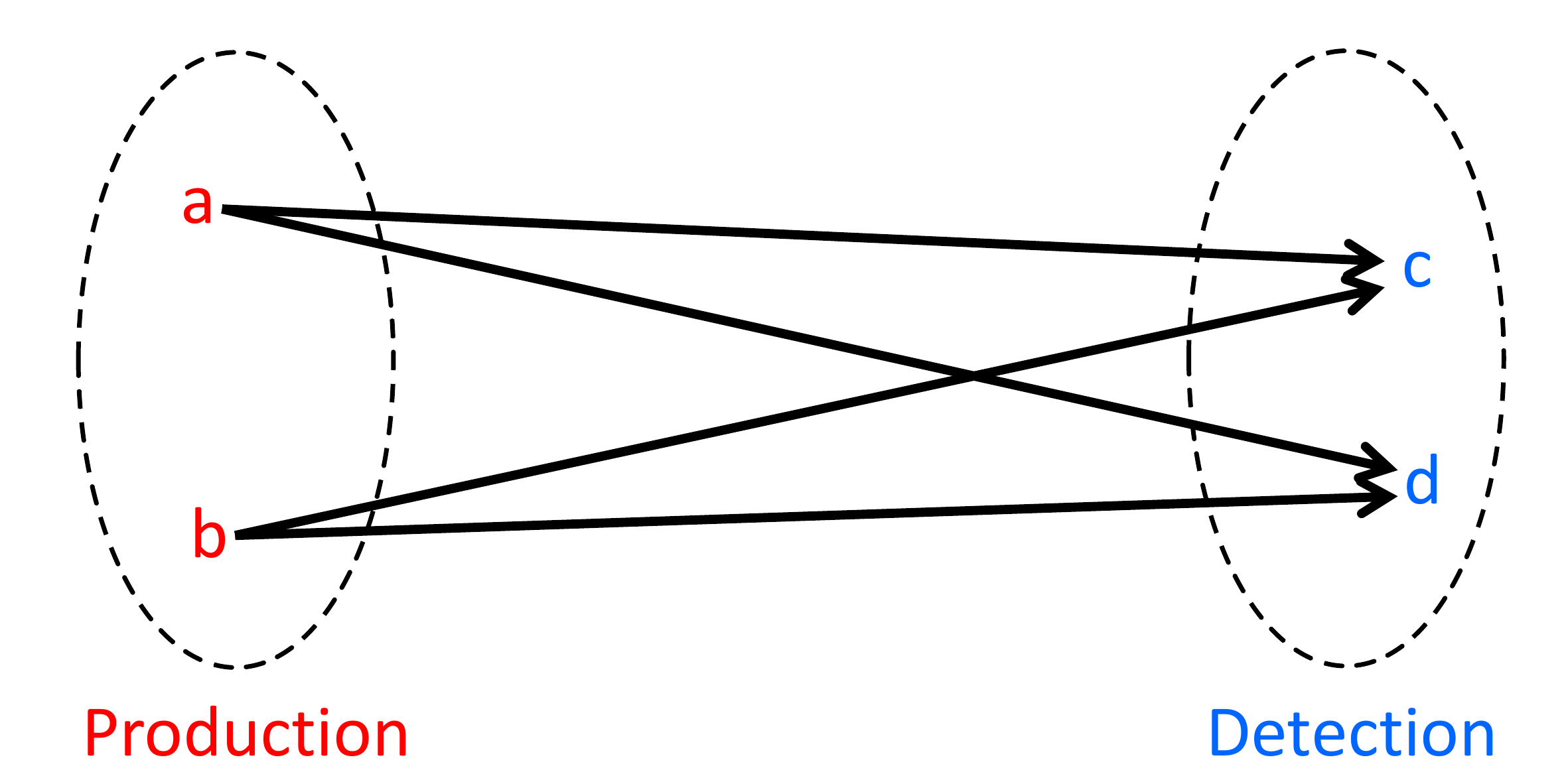}
\caption{Overlapping WPs and the HBT effect. Ambiguity in pairing 
the production and detection processes gives rise to the HBT effect.}
\label{fig:HBT}
\end{figure}

If more than one neutrino can be detected simultaneously, 
then the Hanbury Brown and Twiss (HBT) effect may take place. 
For fermions such as neutrinos, the HBT effect causes the
detected events to ``anti-bunch'' if certain criteria are met 
\cite{Fano1961}. For this to occur, the WPs from production 
points a and b must overlap at detection pionts c and d as 
shown in Fig.~\ref{fig:HBT}. More specifically, the geometric 
condition
\begin{equation}
E_\nu \,r_{ab,\perp}r_{cd,\perp} / D \lesssim 1
\label{eq:hbt}
\end{equation}
must be satisfied and the temporal separation between the two 
detected events must be less than the coherence time of the source.
In Eq.~(\ref{eq:hbt}), $r_{ab,\perp}$ ($r_{cd,\perp}$) is the distance 
between points a and b (c and d) in the direction perpendicular 
to the source-detector axis and $D$ is the source-detector distance.
When the above conditions are realized, the two neutrinos would be 
rendered in the same phase space cell by the detection processes.
Therefore, such joint detection would be suppressed by the Pauli 
exclusion principle, which accounts for the anti-bunching.

Using the parameters given in Table~\ref{tab:overlap factor for sources},
we find that the geometric condition in Eq.~(\ref{eq:hbt}) is only satisfied
for neutrinos from a Galactic supernova. However, even with a megaton 
detector, the expected number of $\bar\nu_e+p\to n+e^+$ events from 
overlapping WPs is $\sim 10^{-14}(\text{kpc}/D)^{2}$, which is
simply too small to show the HBT effect.

\section{Summary}
Based on the evolution of 3D Gaussian WPs, we have estimated the
potential overlap among WPs for accelerator, reactor, solar and supernova
neutrinos. We find that no overlap occurs for the former two sources 
and that the overlap for the latter two does not have measurable
consequences. For analyzing detection of neutrinos from the
above four sources, it is appropriate to treat neutrinos as separate WPs.

\Acknowledgements
This work was supported in part by the U.S. DOE under grant DE-FG02-87ER40328.
C.H.L. gratefully acknowledges the organizers of NuPhys2015 for hospitality and
the Council of Graduate Students at the University of Minnesota for a travel grant.

\end{document}

%% file: econfmacros.tex
\def\beq{\begin{equation}}
\def\eeq#1{\label{#1}\end{equation}}
\def\eeqn{\end{equation}}

\def\beqa{\begin{eqnarray}}
\def\eeqa#1{\label{#1}\end{eqnarray}}
\def\eeqan{\end{eqnarray}}

\let\bar=\overbar

\def\Dslash{\not{\hbox{\kern-4pt $D$}}}
\def\dslash{\not{\hbox{\kern-2pt $\del$}}}

\def\msb{{\bar{\ssstyle M \kern -1pt S}}}

